\documentclass[12pt]{article} 

\usepackage{graphicx} 
\usepackage{mathtools}
\usepackage{amsmath}
\usepackage{multirow}
\usepackage[export]{adjustbox}
\usepackage[left=1in, right=1in, top=1in, bottom=1in]{geometry} 
\usepackage{booktabs}
\usepackage{rotating}
\usepackage{rotfloat}
\usepackage{float}
\usepackage{subfig}
\usepackage{url}
\usepackage{pdfpages}
\usepackage{rotating}
\usepackage[doublespacing]{setspace} 
\usepackage{csvsimple}
\usepackage{array}
\usepackage{wrapfig}
\usepackage[utf8]{inputenc}

\usepackage{tabularx}
\usepackage{float}
\usepackage{verbatim}
\usepackage{booktabs}
\usepackage{rotating}

\usepackage{amsthm}                
\usepackage{amssymb}

\usepackage{multirow}

\usepackage{caption}

\usepackage{amsfonts}
\usepackage{amsmath} 
\usepackage{tikz}
\usetikzlibrary{positioning,shapes.geometric}

\usepackage{xcolor}

\usepackage{setspace}

\usepackage{mathrsfs}

\makeatletter
\newcommand*{\indep}{%
  \mathbin{%
    \mathpalette{\@indep}{}%
  }%
}
\newcommand*{\nindep}{%
  \mathbin{
    \mathpalette{\@indep}{\not}
  }%
}
\newcommand*{\@indep}[2]{%
  \sbox0{$#1\perp\m@th$}
  \sbox2{$#1=$}
  \sbox4{$#1\vcenter{}$}
  \rlap{\copy0}
  \dimen@=\dimexpr\ht2-\ht4-.2pt\relax
  \kern\dimen@
  {#2}%
  \kern\dimen@
  \copy0 
} 
\makeatother

\usepackage{xr-hyper}

\definecolor{forestgreen}{RGB}{34,139,34}
\usepackage{hyperref}
\hypersetup{
    colorlinks=true,
    linkcolor=magenta,
    filecolor=magenta, 
    citecolor=forestgreen,      
    urlcolor=blue
}

\usepackage{tikz}
\usepackage{lscape}
\usepackage{subfig}
\usepackage{float}
\usetikzlibrary{positioning}
\usetikzlibrary{external}
\usepackage{mathtools}
\usepackage{amsmath}
\usepackage{subcaption}
\usepackage{multirow}
\usepackage[export]{adjustbox}
\usepackage{adjustbox}

\usepackage{sectsty}
\sectionfont{\large}

\usepackage[labelsep=endash]{caption}

\usepackage[english]{babel}
\usepackage{blindtext}

\usepackage{layout}

\usepackage{geometry}

\newcolumntype{C}[1]{>{\centering\arraybackslash}p{#1}}

\usepackage{authblk}

\usepackage{afterpage}
\usepackage{lscape}
\usepackage{graphicx}
\usepackage{array}
\usepackage{pdflscape}
\usepackage{longtable}
\usepackage{colortbl}
\usepackage{tabularray}

\usepackage{datetime}
\usepackage{fancyhdr} 
\usepackage{xcolor}

\def\paperversionmajor{9}
\def\paperversionminor{0}

\newtimeformat{24h60m60s}{\twodigit{\THEHOUR}.\twodigit{\THEMINUTE}.\twodigit{\THESECOND}}
\settimeformat{24h60m60s}
\ddmmyyyydate 

\makeatletter
\newcommand*{\addFileDependency}[1]{
  \typeout{(#1)}
  \@addtofilelist{#1}
  \IfFileExists{#1}{}{\typeout{No file #1.}}
}
\makeatother

\newcommand*{\myexternaldocument}[1]{%
    \externaldocument{#1}%
    \addFileDependency{#1.tex}%
    \addFileDependency{#1.aux}%
}

\myexternaldocument{appendix}

\usepackage{amsthm}

\usepackage{hyperref}

\usepackage{array}
\usepackage{pdflscape}
\usepackage{longtable}
\usepackage{rotating}
\usepackage{tabularray}
\usepackage{caption}
\usepackage[backend=biber,style=nature,]{biblatex}
\bibliography{bibliography}

\usepackage{titlesec}

\titleformat{\subsubsection}
  {\normalfont\normalsize\itshape}{\thesubsubsection}{1em}{}

\begin{document}

\title{Using the target trial framework for combining information: external comparator analyses and other applications}

\author[1,2]{Lawson Ung}
\author[1-3]{Miguel A. Hern\'an}
\author[1-4]{Issa J. Dahabreh\thanks{Address for correspondence: Dr. Issa Dahabreh, Department of Epidemiology, Harvard T.H. Chan School of Public Health, Boston, MA 02115; email: \href{mailto:idahabreh@hsph.harvard.edu}{idahabreh@hsph.harvard.edu}; phone: +1 (617) 495‑1000.}}

\affil[1]{CAUSALab, Harvard T.H. Chan School of Public Health, Boston, MA}
\affil[2]{Department of Epidemiology, Harvard T.H. Chan School of Public Health, Boston, MA}
\affil[3]{Department of Biostatistics, Harvard T.H. Chan School of Public Health, Boston, MA}
\affil[4]{Richard A. and Susan F. Smith Center for Outcomes Research, Beth Israel Deaconess Medical Center, Boston, MA}

\maketitle{}
\thispagestyle{empty} 
\clearpage
\vspace*{0.6in}

\thispagestyle{empty}

\begin{abstract}
\noindent
\linespread{1.3}\selectfont
We describe how the target trial framework can be used to plan and report analyses that attempt to answer causal questions by combining information from multiple, diverse sources. Such analyses may involve comparisons of treatments evaluated in different populations, for example when an index trial is combined with other data sources in external comparator analyses, or when extending causal inferences from a randomized trial to a new target population in generalizability and transportability analyses. When planning such analyses, the specification of the target trial supports the explicit definition of the target population with an associated sampling model. We propose this as an additional component for the target trial framework, especially relevant for analyses that combine information, because it influences the choice of eligibility criteria, the specification of the causal model, the choice of causal contrasts, and reasoning about identification strategies. Furthermore, the framework encourages careful mapping of data elements from multiple data sources to a single target trial. This mapping process can highlight potentially irreconcilable misalignments between data sources with respect to specific components of the framework -- for example, in the definitions of eligibility criteria, treatment assignment, and treatment receipt. Such misalignments can arise when attempts to specify a target trial that aligns with a specific data source introduce or worsen misalignments with other proposed data sources. The extent of such misalignments may warrant switching to other data sources, or prospectively obtaining data, to emulate the proposed target trial. We conclude that the target trial framework promotes transparent discussion about the design of and assumptions made in analyses that answer causal questions by combining information from diverse sources.\end{abstract}

\clearpage
\setcounter{page}{1}
\section*{INTRODUCTION}

The target trial framework \cite{hernan2016using, hernan2025} helps guide the planning and conduct of observational analyses to answer questions about the effects of interventions, building on a long tradition of viewing causal inference through the lens of randomized experiments \cite{dorn1953philosophy, wold1954causality, cochran1965planning, cochran1972trial, rubin1974, robins1986, horwitz1987experimental}. The framework calls for specifying the protocol of a hypothetical trial -- the target trial -- that would answer the causal question(s) of interest, and emulating the target trial to the extent possible by mapping the observational data and analysis to its protocol. When studying medical interventions, most target trial emulations share several characteristics. First, the data used for emulation are usually obtained from a single source, such as a prospective cohort study or healthcare database, and are modeled as if drawn from a single underlying population. Second, the target population \cite{robins1987foundations, maldonado2002estimating, rothman2008modern, stuart2018generalizabilitydesignanalysis} is implicitly defined according to the choice of eligibility criteria and the data source used for emulation \cite{westreich2019target, dahabreh2023using}. Third, investigators are usually interested in estimating the effects of treatments that are not randomly assigned, typically using identification strategies that rely on the assumption of no treatment-outcome confounding. 

The target trial framework, however, is agnostic to the types and configurations of data that can be used for emulation. We propose that the framework can be implemented for analyses that attempt to answer causal questions by combining information from diverse sources, such as external comparators for trials  \cite{gehan1974non, pocock1976combination, sacks1982, dempster1983combining, carrigan2019, tan2022, fda2023, hampson2024combining, ung2026combining, ung2026constructing}, and extending inferences from a trial to a target population \cite{cole2010, hernan2016discussionkeiding, stuart2015assessing, buchanan2018generalizing, stuart2018generalizabilitydesignanalysis, dahabreh2019extending_eje, dahabreh2019generalizingbiometrics}. Beyond calling for such analyses to be guided by the specification of a target trial, the framework can address issues particularly salient to combining information for casual inference. These include the need to define a target population of interest alongside a reasonable sampling model for the combined data \cite{dahabreh2021studydesigns, ung2025eje}; highlighting differences between the target trial and each data source with respect to the mapping of protocol components (e.g., when refining causal estimands); and encouraging transparent discourse about less conventional identifiability assumptions that justify combining information. While some analyses have implemented the target trial framework to answer specific research questions by combining information \cite{norvang2022using, polito2024applying, hampson2024combining, arnold2024application, gupta2025quantitative}, here we offer a general methodological elaboration of how the target trial framework can both support these analyses, and highlight complexities when working with multiple sources of data.

\section*{EXTERNAL COMPARATOR ANALYSES}

To motivate ideas, suppose investigators wish to compare the efficacy and safety of different anticoagulation strategies in nonvalvular atrial fibrillation. Specifically, they are interested in comparing the direct thrombin inhibitor dabigatran against other direct acting anticoagulants \cite{connolly2009dabigatran, patel2011rivaroxaban, connolly2011apixaban, giugliano2013edoxaban} in terms of their 24-month risk on stroke and systemic embolism, and on major bleeding events. With no head-to-head trials available or planned, investigators entertain two potential target trials (Tables \ref{tab:tte1} and \ref{tab:tte2}) that use external data sources to examine additional comparator interventions for the RE-LY trial, which compared dabigatran to warfarin \cite{connolly2009dabigatran}. The first emulation compares several anticoagulation strategies by combining the RE-LY trial with electronic health records containing linked prescribing and dispensing information. The second emulation compares the factor Xa inhibitor edoxaban to dabigatran by combining the RE-LY and ENGAGE AF-TIMI 48 \cite{giugliano2013edoxaban} trials.

Throughout, we focus specifically on challenges revealed by the target trial framework when defining estimands and reasoning about their identifiability, which become evident when the target trial protocol is iteratively adapted to accommodate constraints imposed by the available data \cite{hernan2026target}. For statistical aspects related to combining information, we refer readers to the growing literature on the topic \cite{cole2010, westreich2017, dahabreh2019relation, dahabreh2019generalizingbiometrics, ventz2022, li2023improving, dahabreh2023efficient, karlsson2024robust, valancius2024causal, dang2025experiment, ung2026constructing, bartolomeis2026, zhu2026review}, while noting that estimation procedures are only valid to extent that the planning and conduct of such analyses can support identification of causal quantities of interest.

\subsection*{Target population and sampling} 

\subsubsection*{Considering the target population and sampling in the target trial framework} When combining information, a primary concern is explicitly characterizing the relationships between the data sources and how they relate to the target population one wishes to learn about. We propose the specification of the target population and sampling model as a useful addition to the target trial framework when combining data because they underlie the specification of components relevant to the causal estimand(s) and the reasoning about their identifiability. Most analyses that combine information from diverse sources implicitly assume a stratified sampling scheme under which data sources are modeled as having been obtained, with known or unknown sampling probabilities, from strata of a hypothetical superpopulation of treatment-eligible individuals \cite{robins1987foundations, robins2002covariance, dahabreh2021studydesigns, dahabreh2024invited}. 

To specify the target population and a reasonable sampling model in our working example, consider that the collection of data from the already-completed RE-LY trial has been conducted separately to the collection of external data in both target trial emulations presented. This is typical of ``non-nested'' sampling models \cite{dahabreh2021studydesigns}. Data from each of these populations are modeled as independent simple random samples from strata of an underlying infinite population of all treatment-eligible individuals; strata are defined by participation in the index RE-LY trial, without assuming that the strata are governed by the same distribution. The sampling fraction for each stratum is unknown, and therefore the ``pooled'' distribution induced by combining data from these populations may not fully represent of the population comprising all treatment-eligible individuals, or any reasonable ``real-world'' population for that matter. In keeping with what is often implicitly assumed in both trials and observational analyses, investigators here explicitly designate the target population as comprising individuals who would be invited and would agree to participate in the RE-LY trial. An inherent tradeoff in this designation is that the target population would not allow investigators to learn about the effect of anticoagulation strategies under non-experimental, ``real-world'' settings. Learning about such populations would require other target trial specifications, for example those consistent with generalizability and transportability analyses where data requirements will differ (see Supplement). 

\subsubsection*{Challenges in specifying the target population when combining information}

Defining the target population of interest is challenging when the definition and ascertainment of eligibility criteria across different sources are misaligned. In most applications, investigators will specify the target trial's eligibility criteria according to the index trial, with the goal of using the external data to construct treatment groups within the target population (sometimes this has been informally described as emulating ``half'' or some portion of a target trial \cite{gupta2025quantitative, hernan2025emulatinghalf, dickerman2026target}). Whether it is feasible to use the external data for this purpose requires careful consideration. In the first proposed target trial emulation, the eligibility criteria of the target trial was specified according to the index RE-LY trial; as a result, one would require justifications for using observational electronic health record data in which covariates are unlikely to fully align with the extensive screening investigations that typically determine trial eligibility. Conversely, specifying the target trial with pragmatic eligibility criteria (e.g., by waiving certain screening procedures), in an effort to align more closely with the external data, may not fully align with how participants were enrolled in the RE-LY trial. 

When the index trial data or external data cannot be satisfactorily mapped to the eligibility criteria of the target trial, investigators may be compelled to search for more compatible sources of data. In the second proposed target trial emulation, investigators may be more confident that there is sufficient alignment between definitions and the ascertainment of eligibility criteria because the data are derived from RE-LY and ENGAGE, which were similar landmark trials. Nevertheless, one must still be thoughtful about handling residual misalignments in eligibility criteria. For example, the ENGAGE trial enrolled participants with a CHADS$_2$ \cite{gage2001chads} score of $\geq 2$, while the RE-LY trial also included participants with CHADS$_2$ scores of 0 and 1. Aligning these trials would require restricting RE-LY to individuals with a baseline CHADS$_2$ score of $\geq 2$, at the cost of narrowing the target population to one at higher baseline risk of stroke than in the first target trial emulation.

Furthermore, the application of different procedures for selecting participants (or observations) across data sources can obscure, and sometimes shift, the target population. A common approach to construct external comparator groups is selecting only known initiators of the treatments of interest, to the systematic exclusion of individuals who may have initiated other treatment strategies (including non-treatment). Such approaches may be motivated in part by the desire to mirror the index trial, which may not have randomized participants to non-treatment (as was the case in RE-LY), and the belief that comparing newly initiated treatments (e.g. as part of ``new-user'' analyses) may be less prone to treatment-outcome confounding \cite{ray2003, schneeweiss2007increasing, lund2015active, yoshida2015active, schneeweiss2019advanced}. However, in the target trial, as in most actual trials, known treatment initiation \textit{cannot}, by definition, be an eligibility criterion. Furthermore, such an approach would result in a misalignment between populations: the trial-eligible individuals in the index trial, and the initiators-only subset of trial-eligible individuals from the external data. Such ambiguity in defining the target population may complicate the analyses, for example when specifying causal estimands and choosing the populations to whom causal effects will be marginalized, as well as reasoning about population-specific identifiability conditions \cite{huitfeldt2016comparative}. These challenges when constructing external comparators are part of a pervasive pattern of selecting (conditioning) on treatment initiation in the epidemiologic literature, which almost always results in a shift in the target population and in some cases can lead to bias \cite{huitfeldt2016comparative, chiu2020effect}. 

\subsection*{Defining causal estimands} 

A central goal of the target trial framework is to specify components of the trial to define and refine causal estimands without necessarily resorting to mathematical formalism \cite{rubin1974, robins1986, hernan2016using}. In our target trial emulation, investigators are interested in the intention-to-treat and per-protocol effects of different anticoagulation strategies within the population underlying the index RE-LY trial. These causal estimands refer to the effect of being assigned a particular anticoagulation strategy, and of being assigned and following a particular anticoagulation strategy, respectively \cite{hernan2017perprotocol, dahabreh2025roleassignment}. 

When defining causal estimands that involve intervention on assignment, and joint intervention on assignment and treatment \cite{dahabreh2025roleassignment}, the target trial framework helps investigators reason through potentially irreconcilable misalignments in how assignment and treatment are defined across the data sources. For example, when mapping interventions between the index trial and observational data, one usually makes the implicit assumption that interventions on assignment and treatment receipt are sufficiently similar to their observational analogs, treatment prescription and dispensation, respectively. Such assumptions -- sometimes described as assumptions of \textit{treatment variation irrelevance} \cite{vanderweele2009further} -- may be called into question when the procedural implementation of assignment and treatment differs between sources, e.g., when assignment is blinded in one source but not the other. Moreover, when no observational analogs of assignment exist in the data, emulating a target trial by combining information will not be amenable to answering questions about the effect of assignment, and possibly per-protocol effects, unless further strong assumptions are made \cite{dahabreh2025roleassignment}.

\subsection*{Identifiability assumptions} 

Implementing the target trial framework requires investigators to reason explicitly about the conditions needed for the causal estimands of interest to be expressed as functions of the observed data distribution, in the context of an appropriate causal model for the populations under study. Causal models that involve combining information often need to be more expansive than those of single-source emulations because they require reasoning about the relationships within \textit{and} between the underlying populations at hand \cite{pearl2011, bareinboim2016causalfusion, dahabreh2019identification, dahabreh2020benchmarking}, and may involve methods to adjust for both pre- and post-baseline sources of bias. 

A primary concern when combining information is whether there exist \textit{trial engagement effects} or other source effects -- actions, activities, and procedures that accompany trial participation, and that operate though paths that do not involve the intervention under examination \cite{dahabreh2019extending_eje, dahabreh2019identification, dahabreh2020transportingStatMed, ung2025generalizing}. In our example, investigators must be willing to assume that participating in the RE-LY trial did not entail outcome-relevant Hawthorne effects \cite{landsberger1958, sedgwick2015understanding}, behavioral modifications, nor standards of care or follow-up that were not observed within the population underlying the external data (either the electronic health records in the first trial emulation, or the ENGAGE trial in the second). If met, the condition of absent trial engagement effects allows the effect of the intervention to be identified regardless of the setting, trial or non-trial, in which it was administered. On the other hand, the presence of trial engagement effects can substantially complicate identification, and in some cases mean that the effect of the intervention cannot be identified, or cannot be disentangled from the effect of trial participation itself \cite{dahabreh2022adherence, ung2025generalizing}.

Furthermore, the causal model requires assumptions about whether features of the population underlying any specific data source can be used to learn about other populations, usually in terms of their distribution of covariates, treatments, and outcomes. Here, it is assumed that participants in the RE-LY trial and participants in the external data are conditionally exchangeable within levels of measured covariates \cite{pearl2014, bareinboim2016causalfusion, stuart2018generalizabilitydesignanalysis, dahabreh2019generalizingbiometrics, dahabreh2020transportingStatMed}. This conditional exchangeability assumption, commonly invoked for generalizability or transportability analyses \cite{pearl2011, bareinboim2012transportability, dahabreh2021studydesigns}, is distinct from exchangeability over treatment. This is because the former involves the assumption that some features of the distribution of potential outcomes under intervention on treatment are independent of trial participation status, rather than of treatment itself \cite{pearl2011}. 

Finally, emulating a target trial using multiple sources of data often requires subtle modifications of the identifiability assumptions necessary for single-source emulations. For example, one requires consistency of potential outcomes \cite{rubin1986ifs, hernan2008a, vanderWeele2009, hernan2016does} to hold within \textit{and} across data sources. Investigators also require some form of conditional exchangeability over the intervention of interest within each data source, both pre-baseline (for intention-to-treat effects) and post-baseline (for per-protocol effects). For these assumptions to hold in the external data, one may require an expanded set of covariates not captured in the index trial.

\section*{WHAT THE TARGET TRIAL FRAMEWORK OFFERS FOR COMBINING INFORMATION }

Our running example of external comparator analyses, as well as the additional examples presented in the Supplement, show how the target trial framework can support the planning and conduct of analyses that combine information from diverse sources to answer causal questions. Taken together, these examples highlight three main contributions of the target trial framework to analyses that combine information.

\subsection*{Treating the target population as a key component of the estimand}

When combining information, there is a need to clearly articulate the target population of interest to motivate scientific decision making as the target trial and its emulation are specified \cite{dahabreh2021studydesigns, dahabreh2023using, ung2025eje}. Leaving the target population and sampling model implicit can allow serious deficiencies in the specification of the eligibility criteria, treatment strategies, causal models and estimands, identifiability conditions, and estimation procedures to go unnoticed. For example, constructing an external comparator group by sampling the target population on the basis of known treatment initiation, to the systematic exclusion of non-initiators, can create challenges in characterizing the target population and in some cases induce bias in the analyses when deciding the populations to whom effects will be marginalized \cite{chiu2022selection}. The importance of explicitly describing the target population and sampling model, particularly when using multiple sources of data for trial emulation, compelled its addition as a distinct component of the target trial framework.

\subsection*{Highlighting and reconciling misalignments}

The target trial framework encourages careful consideration of whether the available data sources can be sufficiently aligned to the specified target trial, and whether the underlying population satisfies the identifiability conditions that justify combining information. Using data from multiple sources poses a unique challenge in that specifying a target trial that closely aligns with a specific index study, as is most often done in practice, may render the additional external data incompatible with the target trial. Conversely, specifying an ``emulatable'' target trial that accommodates the use of the external data, for example by incorporating pragmatic eligibility criteria, may induce misalignments of the target trial with the index study. These misalignments often manifest when defining the target population and specifying causal contrasts that involve intervention on assignment and possibly also on treatment receipt. These issues extend beyond the target trial's well-recognized role in revealing biases created by the failure to anchor the start of follow-up to eligibility assessment and treatment assignment, which could itself be viewed as an issue of misalignment \cite{hernan2016a, hernan2025structural}.

Potential approaches to address misalignments when combining information include (1) designing the index trial with the plan to incorporate external data in the analyses, e.g., by defining trial eligibility criteria with pragmatic elements that could be mirrored, or reasonably approximated, in the external data; (2) designing the collection of external data to align with a planned or existing trial, for example, by enrolling a prospective cohort defined by eligibility criteria that are assessed in the same manner as in a trial, and whose participants are exposed to similar background care and monitoring intensity; or (3) searching for more suitable sources of existing data, for instance by combining information using clinical trials where the eligibility criteria, and the meaning of assignment and treatment receipt can be unambiguously harmonized to the same target trial, and where important identifiability assumptions may be more plausible. 

Whether any of these solutions are appropriate to inform clinical, regulatory, and policy decisions will always require context-specific judgments. The first two approaches should invite questions concerning why a randomized trial directly comparing all the treatments of interest was not conducted in the first instance. Still, in some cases it may be necessary to adopt one of these two approaches on the basis of ethical constraints (e.g., limitations in obtaining consent, changes in equipoise); research economy considerations (e.g., challenges in participant recruitment and retainment, commonly encountered in rare disease and cancer settings \cite{kempf2018challenges, rahman2023external}); and when trials are impractical owing to the nature of the intervention itself (e.g., when the results of medical device trials become obsolete in view of rapid and successive refinements to the technology \cite{pocock2014current, holmes2016overcoming}). The third potential solution may sometimes be a practical alternative, but the range of treatments that can be studied using trial data is limited by the availability of suitable high-quality trial evidence.

\subsection*{Reasoning about assumptions that allow combining information}

The target trial framework offers a structured approach to communicate the assumptions necessary to answer causal questions when combining information. Recently updated guidance on target trial implementations \cite{hernan2025, cashin2025transparent} call for the explicit articulation of identifiability conditions that allow causal estimands to be expressed as observed data quantities. Though practical applications of combining information often rely on novel identification strategies that require explicit reasoning about the relationships within elements of the joint data distribution \cite{pocock1976combination, cole2010, stuart2011, stuart2018generalizabilitydesignanalysis, dahabreh2019generalizingbiometrics, dahabreh2020transportingStatMed, dahabreh2021studydesigns, dahabreh2023using, ung2026combining, ung2026constructing}, and require tailored methods to adjust for pre- and post-baseline sources of bias, seldom are these conditions openly communicated.

Causal models that can accommodate multiple, diverse data sources not only demand conditions necessary for valid inference in single-source studies, but also those that justify combining information. The latter calls for reasoning about the extent to which trial engagement effects are present; whether the populations underlying the index trial an external data are conditionally exchangeable; and whether background knowledge will allow any of these assumptions to be relaxed or empirically tested using the data \cite{pocock1976combination, pearl2011, bareinboim2012transportability, bareinboim2016causalfusion, dahabreh2019generalizingbiometrics, dahabreh2019identification, dahabreh2020transportingStatMed, dang2025experiment, ung2025generalizing, ung2026combining, ung2026constructing}. The strength of these assumptions and the stringent data requirements to approximately satisfy them suggest that external control analyses should in most settings be viewed as an adjunct to, rather than replacement, for clinical trials.

\section*{CONCLUSION}

We described methodological elaborations of the target trial framework that may be useful when planing analyses that combine information from diverse sources. The framework can help clarify causal estimands and target populations of interest, but it requires explicit reasoning about misalignments between the different data sources and their underlying populations, and transparent discussions about the causal structure and conditions needed to identify causal quantities of interest. The value of using the target trial framework to plan such analyses will likely become more evident as methods mature, encouraging investigators to openly communicate their assumptions and analytical choices.

\section*{TARGET TRIAL EMULATION TABLES}

The following two target trial emulation tables are referenced in the main text. Both specify identification strategies that involve baseline covariates, including age; sex; race; body mass index; type of atrial fibrillation (paroxysmal, permanent, or persistent); CHADS$_2$ score consisting of congestive cardiac failure, hypertension, age $\geq 75$, diabetes mellitus, and history of stroke and transient ischemic attack (TIA) \cite{gage2001chads}; comorbidities including prior stroke or TIA; coronary artery disease; previous myocardial infarction; heart failure; diabetes mellitus; hypertension; and medication use at baseline including aspirin, angiotensin converting enzyme inhibitors, angiotensin receptor blockers, beta-blockers, lipid-lowering agents, vitamin K antagonists, and anti-arrhythmics. Post-baseline covariates include CHADS$_2$ score, comorbidities, and medications evaluated every three months after the start of follow-up. 

\begin{landscape}
\begin{table}
    \caption{Emulating a target trial by combining the RE-LY trial with electronic health records to evaluate anticoagulation treatment strategies in terms of (1) stroke and systemic embolism; and (2) major bleeding events at 24 months.}\vspace{-0.2in}\centering
    \includegraphics[width=0.95\linewidth]{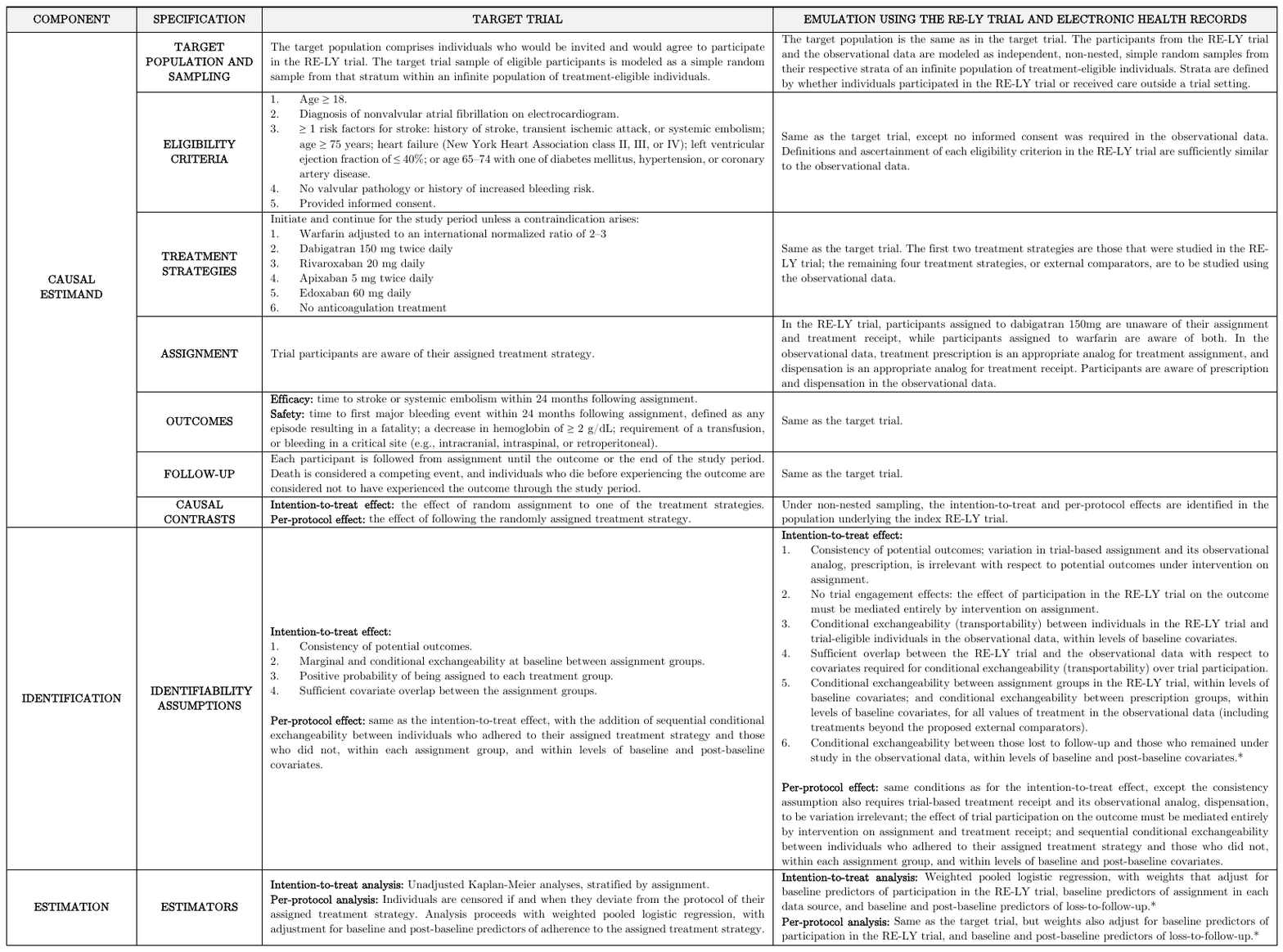}
    \label{tab:tte1} \vspace{-0.3in}
    \begin{flushleft}\footnotesize
        $^*$There was near-complete follow-up in RE-LY; adjustment for censoring due to loss to follow-up may be required in the observational data.
    \end{flushleft}
\end{table}
\end{landscape}

\newpage
\begin{landscape}
\begin{table}
    \caption{Emulating a target trial by combining the RE-LY and ENGAGE AF-TIMI 48 trials to compare dabigatran 150mg with edoxaban 60mg in terms of (1) stroke and systemic embolism; and (2) major bleeding events at 24 months. There was near-complete follow-up in both trials.}\centering \vspace{-0.8in}
    \includegraphics[width=1\linewidth]{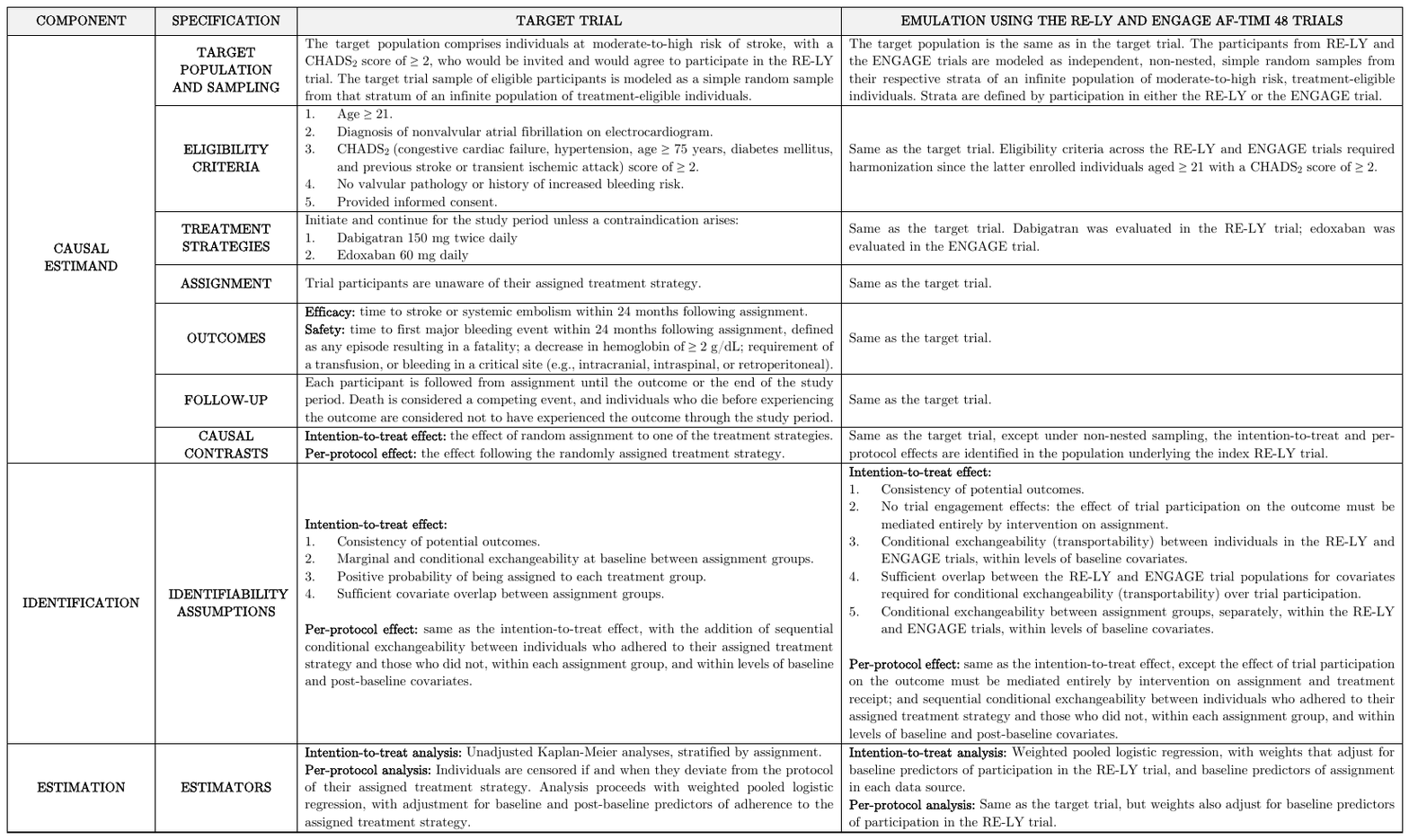}
    \label{tab:tte2}
\end{table}
\end{landscape}

\clearpage
\renewcommand{\refname}{REFERENCES}
\printbibliography


\pagestyle{fancy}
\fancyhf{} 
\renewcommand{\headrulewidth}{0pt} 

\fancyfoot[C]{
    \textcolor{gray}{\tiny 
        combining\_target\_trial, 
        Date: \today\ \currenttime\ \ 
        Revision: \paperversionmajor.\paperversionminor
    }
}

\end{document}